\documentclass[lettersize,journal]{IEEEtran}
\usepackage{amsmath,amsfonts,amssymb}
\usepackage{algorithmic}
\usepackage{algorithm}
\usepackage{array}
\usepackage{booktabs}
\usepackage[T1]{fontenc}
\usepackage[rgb,x11names]{xcolor}
\usepackage{hyperref}
\usepackage{mathrsfs}
\usepackage{pgfplots}
\usepackage{textcomp}
\usepackage{stfloats}
\usepackage{url}
\usepackage{verbatim}
\usepackage{graphicx}
\usepackage{cite}
\usepackage{framed,multirow}
\usepackage{float}

\def\BibTeX{{\rm B\kern-.05em{\sc i\kern-.025em b}\kern-.08em
    T\kern-.1667em\lower.7ex\hbox{E}\kern-.125emX}}
    
\begin{document}
\title{
InstantGroup: Instant Template Generation for Scalable Group of Brain MRI Registration}%

\author{
Ziyi He and Albert C. S. Chung
\thanks{Ziyi He and Albert C. S. Chung are with the Department of Computer Science and Engineering, The Hong Kong University of Science and Technology, Clear Water Bay, Hong Kong. (e-mail: zheaj@connect.ust.hk, achung@cse.ust.hk). 
This work was supported by the Hong Kong Research Grants Council under Grant 16214521.}
}

\maketitle
\begin{abstract}
Template generation is a critical step in groupwise image registration, which involves aligning a group of subjects into a common space. While existing methods can generate high-quality template images, they often incur substantial time costs or are limited by fixed group scales. In this paper, we present InstantGroup, an efficient groupwise template generation framework based on variational autoencoder (VAE) models that leverage latent representations' arithmetic properties, enabling scalability to groups of any size.
InstantGroup features a Dual VAE backbone with shared-weight twin networks to handle pairs of inputs and incorporates a Displacement Inversion Module (DIM) to maintain template unbiasedness and a Subject-Template Alignment Module (STAM) to improve template quality and registration accuracy. Experiments on 3D brain MRI scans from the OASIS and ADNI datasets reveal that InstantGroup dramatically reduces runtime, generating templates within seconds for various group sizes while maintaining superior performance compared to state-of-the-art baselines on quantitative metrics, including unbiasedness and registration accuracy.

%%%%
\end{abstract}
\begin{IEEEkeywords}
Deep Generative Models, Template Generation, Groupwise Registration, Brain MRI images
\end{IEEEkeywords}

\section{Introduction}
\label{sec:introduction}
\begin{figure*}[!t]
\centerline{\includegraphics[width=1.55\columnwidth]{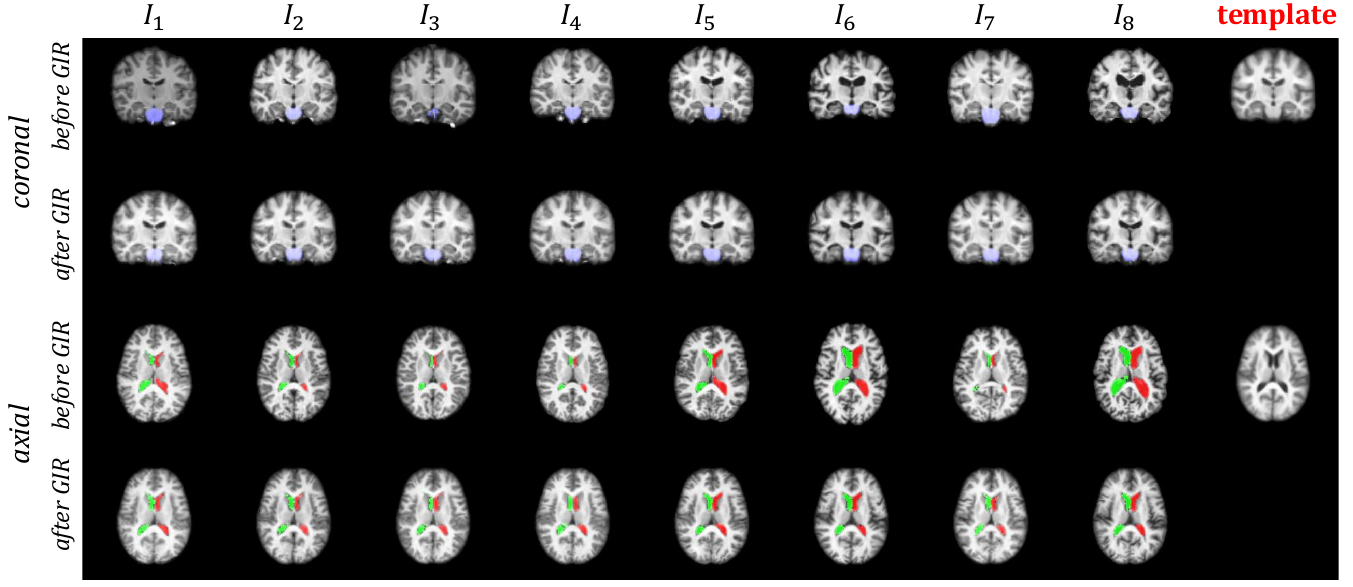}}
\vspace{-0.4cm}
\caption{Visualization of GIR on a sample group of brain MRIs with the template generated by InstantGroup. The figure shows coronal and axial views. The MRI scans pre-GIR and post-GIR are displayed in rows 1, 3, and 2, 4, respectively. Blue, green, and red masks denote the brain stem and left and right ventricles. The generated template is shown in the rightmost column in rows 1, 3.}
\label{intro}
\vspace{-0.3cm}
\end{figure*}

\IEEEPARstart{A}{ligning} anatomical structures into a common reference space is crucial for identifying and measuring structural differences and changes indicative of various conditions or diseases while minimizing individual differences \cite{fonov2011unbiased, jenkinson2002improved}. 
In medical image analysis, this alignment facilitates enhanced diagnosis and effective treatment planning by providing a standardized way to compare anatomical features, thus reducing the variability that can obscure meaningful differences \cite{avants2008symmetric, ashburner2007fast}. For instance, subtle changes in brain structure due to neurodegenerative diseases like Alzheimer's can be more readily identified when all subjects are registered to a common template \cite{mueller2005ways}.
The procedure involving aligning a set of images into a common space is groupwise image registration (GIR), a fundamental task in computer vision, medical image analysis, and time-series signals \cite{joshi2004unbiased}. Unlike pairwise registration that registers one image to another, GIR simultaneously considers multiple images, aiming to find transformations that best map each group image to the common template (an example is shown in Figure \ref{intro} with an explicit template).
Moreover, GIR supports advanced statistical analysis and machine learning applications. Reducing inter-subject variability and increasing statistical significance makes it possible to detect more negligible effects and relationships in the data \cite{friston1995spatial}, leading to more reliable and generalizable predictive models for diagnostic and prognostic purposes \cite{cheplygina2019not}.

However, the common group space is implicit and not readily available most of the time, which presents a significant challenge, as the choice of the template greatly influences registration accuracy and bias. Performing pairwise registration or selecting a subject from the target group or utilizing a general atlas built by other population as the template can introduce uncertainties and bias, resulting in a deviated deformed group losing initial mutual information \cite{joshi2004unbiased, fonov2011unbiased}. 
While some groupwise registration methods do not rely on the template and perform direct groupwise registration \cite{he2020unsupervised, agier2020hubless, ying2014hierarchical}, constructing an explicit template image has distinct advantages. 
A group-tailored template is the representative image that can serve as the average feature of the group. The template can be used as the coordinate system to track variations in longitudinal data analysis \cite{ding2022aladdin}. In addition, group-tailored templates usually have shorter distances to each subject of the group than a general static atlas, which leads to better registration performance.

Classic template construction algorithms such as iterative optimization methods \cite{joshi2004unbiased, wu2011sharpmean, wang2005construction} suffer from high computational costs, especially for 3D volumetric data. 
Some works \cite{dong2019fast, sabuncu2009image} committed to utilizing hierarchical structures that divide large groups into smaller ones. These methods can deal with significant anatomical variants and improve processing efficiency, but they still face overall runtime and scalability limitations. 
Recent advances in learning-based approaches \cite{ mok2020large, mok2021conditional, hoopes2021hypermorph} offer promising solutions by leveraging neural networks to learn optimal transformations and generate high-quality templates efficiently, making GIR more practical for large-scale and real-time applications \cite{dalca2019learning, dey2021generative, ding2022aladdin}.
% learning-based methods
Existing learning-based methods can be categorized into two approaches. The first approach learns parameters for registration and constructs the template image during the training \cite{dalca2019learning, dey2021generative, ding2022aladdin}. 
Despite their effectiveness, these methods can be regarded as neural-network-based optimization; the templates are optimized for each group of images, which is a significant limitation. Another research direction in learning-based GIR focuses on predicting the deformation fields directly\cite{van2020deep, he2020unsupervised} bypassing the need for iterative optimization to achieve faster registration but often constrained to fixed group settings.

Given these challenges, there is a core need for efficient and scalable learning-based GIR methods that can handle varying group sizes without compromising performance. 
We propose to employ a Variational Autoencoder (VAE) \cite{kingma2013auto, rezende2014stochastic} with the assumption of the normal prior distribution of the latent space to perform latent space exploration, including interpolation and arithmetics on the latent vectors \cite{berthelot2018understanding, van2017neural} to generate templates for MRI groups of arbitrary sizes instantly. The assumption of a normal prior aligns with the nature of the brain anatomy, which exhibits smooth and continuous variations across the population, especially for single-modality modeling.
Hence, we reshape and decompose the groupwise template generation into three steps: acquiring the latent representation of each group subject, performing vector averaging, and reconstructing the template from the mean latent vector. This framework ensures stable and unbiased template generation with significantly reduced runtime. Based on it, we introduce the dual VAE backbone inspired by Siamese networks \cite{bromley1993signature, koch2015siamese, varior2016gated} to deal with paired inputs simultaneously using the VAEs with shared weights and get the template. By comparing the displacement fields from the paired symmetric paths to the template, we significantly improve groupwise registration performance. 
To summarize, our contribution can be concluded as follows:
\begin{itemize}
	\item We are the first to utilize the latent representation of Dual VAE with symmetric architectures to generate group-tailored templates efficiently for groupwise MRI image registration. The approach is scalable to problems of any scale with few additional costs;
	\item We propose the Displacement Inversion Module (DIM) to ensure the generated template is unbiased regarding the local deformation fields from all group subjects. We propose the Subject-Template Alignment Module (STAM) to compare images in the input space and improve the generated template;
	\item Experiments on two datasets of 3D brain MRI scans demonstrate that our method can achieve state-of-the-art performance in terms of unbiasedness and accuracy compared with baselines in a much shorter runtime ranging from hours to seconds.
\end{itemize}

\section{Related work}
\label{sec:related}
\subsection{Pairwise Image Registration}
Image registration generally involves aligning the moving image to the reference (fixed) image to minimize dissimilarity between them. Registration is challenging, especially for brains, due to anatomical complexity and significant inter- or intra-subject variations.   
Before deep learning-based image registration methods, traditional optimization-based methods involved the selection of feature space, similarity measurement, the transformation type (local or global, rigid or non-rigid), and optimization strategy \cite{oliveira2014medical}. Rigid or affine transformations are mainly adopted in the registration of solid structures like bones or the pre-alignment before non-rigid registration \cite{andreetto2004frequency,auer2005automatic}. For most scenarios involving deformable subjects and significant anatomical inconsistency, non-rigid transformation, mainly free-form deformation models, are used \cite{schnabel2001generic,periaswamy2003elastic}. 
 
Fueled by the thriving of deep learning technology, many works replace the erroneous iterative optimization with supervised/unsupervised learning or reinforcement learning models using neural networks \cite{chen2021learning,mok2022affine,zhao2019unsupervised, hu2021end, liao2017artificial}. VoxelMorph \cite{balakrishnan2018unsupervised} utilizes the U-Net and the spatial transformer network to predict the deformation fields. Later, the diffeomorphic transformation was incorporated into the unsupervised framework \cite{dalca2018unsupervised} by predicting and integrating the velocity field to get the deformation field. To handle large deformation scenes, Mok and Chung \cite{mok2020large} proposed a Laplacian pyramid network to solve the registration problem in a coarse-to-fine fashion. In many deep learning-based GIR methods, including the proposed method, diffeomorphic registration models are incorporated into the whole framework and generate the template. 
\vspace{-0.1cm}
\begin{figure*}[!t]
    \centering
    \includegraphics[width=1.95\columnwidth]{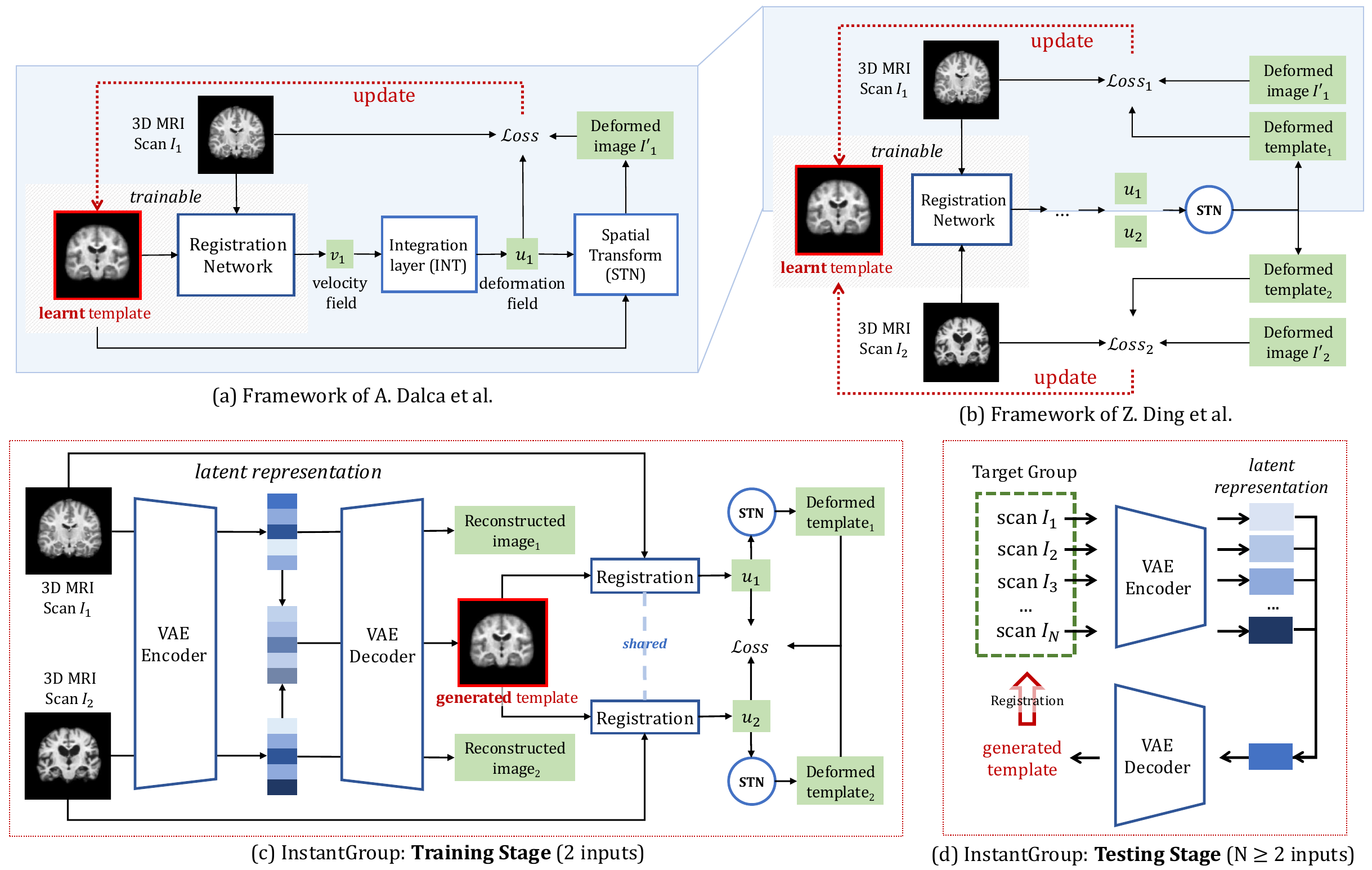}
    \vspace{-0.2cm}
    \caption{Paradigm of baseline (a), (b), and the proposed InstantGroup (c), (d). (a) The baseline \cite{dalca2019learning} for building deformable templates (unconditional version displayed). (b) The baseline \cite{ding2022aladdin}, which can be regarded as a doubled version of (a) with two inputs, incorporates pairwise image similarity losses in both atlas and image space to improve accuracy. Both baselines update the template tensor with each test group subject during the training stage.
    In contrast, the proposed method (c) takes a dual VAE backbone to decode the template from aggregated latent representations of paired inputs. The structure predicts templates instead of updating the template during training like (a) or (b). (d) The testing stage of InstantGroup, different from its training stage, is able to scale to any group size (more than 2) by encoding subjects and averaging their latent vectors to generate the template.}
    \label{fig:overview}
    \vspace{-0.2cm}
\end{figure*}

\subsection{Learning-based groupwise image registration}
Deep learning-based pairwise registration has significantly impacted the norm of GIR.  
Ahmad et al. \cite{ahmad2019deep} used deep neural networks to estimate the large initial deformations in the multi-level graph coarsening framework, improving the convergence speed. Dalca et al. \cite{dalca2019learning} present a probabilistic model with diffeomorphic transforms for jointly template construction and registration, supporting conditional template generation given attributes. Dey et al. \cite{dey2021generative} added a discriminator to the framework for more authentic templates. 
Aladdin \cite{ding2022aladdin} introduced pairwise image losses with inputs in pairs and evaluated in image space via Atlas-as-a-bridge, improving overlap degree but lacking explicit template unbiasedness evaluation. These methods require optimization of the target group, limiting flexibility and increasing time consumption, especially with large datasets or varying group sizes.

Another approach develops models to predict for different groups.
He et al. \cite{he2020unsupervised} proposed an unsupervised learning scheme to predict deformation fields to warp subjects without explicit templates. Semantic information is used \cite{he2021learning} in a segmentation-assisted GAN to improve GIR performance. GroupMorph \cite{van2020deep} learns diffeomorphic velocity field distribution with the variational GIR method. 
Wang et al. \cite{wang2024groupwise} proposed a framework to isolate respiratory and cardiac motions with implicit templates and preserved diffusion contrast textures with a tensor-embedded branch. PCA-relax \cite{zhang2024deep} is another groupwise registration method for cardiac motion correction with a PCA loss from the PCA decomposition of warped group subjects. 
GMM-CoRegNet \cite{li2024gmm} is a weakly supervised framework for multimodal groupwise registration that models anatomical structures as prior information and proposes a similarity measure to register group subjects to the assigned template.
These methods have short runtime due to end-to-end learning but are restricted to fixed group sizes, where networks trained on a specific group size could not generalize well to groups of different sizes.
In this paper, we present InstantGroup, a novel GIR framework that combines the efficiency of learning-based models with the scalability of optimization-based models. InstantGroup can be applied to any new group of any size without additional training or optimization.
\vspace{-0.1cm}

\subsection{Deep generative models}
GIR usually involves generating a center image representing the target group. 
In the past decade of research, various deep generative models which learn the data representation and synthesize realistic images have shown great potential in high-quality image generation. Generative Adversarial Nets (GAN) \cite{goodfellow2020generative} is proposed to model the generation as an adversarial process where the generative model learns the data distribution and the discriminative model evaluates the authenticity of the generated image. Later, conditional GANs \cite{mirza2014conditional} model the conditional distribution to generate specified samples given auxiliary conditions. CycleGANs \cite{zhu2017unpaired} is another landmark for translating an image from one domain to another without the need for training data in pairs. Research on GANs and variants \cite{zhang2017stackgan, karras2017progressive} remains active today, and many are applied to medical modalities \cite{karras2019style, denton2015deep, mao2017least, reed2016generative, kang2019cycle, armanious2019retrospective} including MRI and CT.  

Although GANs are known for efficiency and high-quality results, they usually suffer from unstable training and limited sample diversity. On the other hand, Variational Autoencoders (VAEs) \cite{kingma2013auto} learn the data likelihood distribution explicitly. VAEs are comprised of two components: the encoder that compresses data into a low-dimensional latent space and the decoder that reconstructs data from the latent input. 
More recently, VQ-VAE \cite{van2017neural} trains the autoencoder with a discrete latent space and maintains a codebook of latent vectors, allowing it to adapt to natural modalities and learn complex data distributions. Other related works for medical image analysis \cite{shakeri2016deep, biffi2018learning,elbattah2021variational, sandfort2021use, dorent2019hetero, zimmerer2018context} illustrate the superior performance of VAEs in representation learning.
Autoencoders have demonstrated their ability to interpolate latent vectors in many studies \cite{roberts2018hierarchical, ha2017neural}, suggesting that data can be semantically distributed in latent space \cite{berthelot2018understanding}. While the existing VAE algorithms cannot meet the requirements of atlas construction of groupwise registration, we propose additional regularization on VAEs to encourage the interpolated outputs to be closer to brain MRI scans and correlate the latent interpolation with registration distances. 

\begin{figure*}[!t]
\centerline{\includegraphics[width=2\columnwidth]{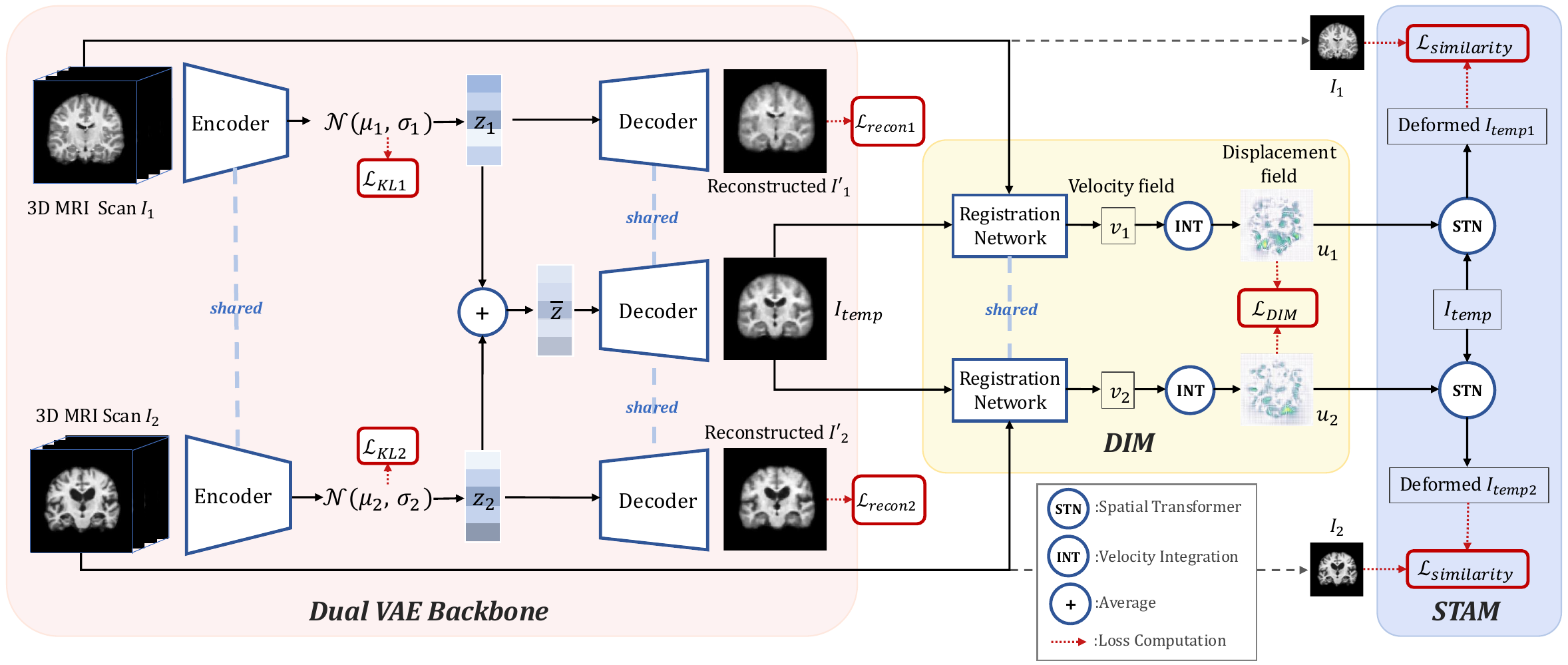}}
\vspace{-0.2cm}
\caption{Overview. The proposed method, InstantGroup, leverages a Dual VAE backbone for efficient and scalable high-quality template generation with DIM (Displacement Inversion Module) and STAM (Subject-Template Alignment Module). 
The Dual VAE backbone is a symmetric dual-path VAE architecture consisting of symmetrical upper and down parts with shared-weights encoder and decoder, allowing for the computation of the average latent vector $\bar{z}$ and the generation of the template $I_{temp}$ for each pair of input MRI scans. 
DIM ensures unbiased template generation at the level of registration distance with a diffeomorphic registration network applied to predict velocity fields $v$, integrated into deformation fields $\phi$ to deform $I_{temp}$. 
STAM improves template quality by measuring the similarity between the deformed template and each input scan.
}
\vspace{-0.2cm}
\label{fig:framework}
\end{figure*}

Besides GANs and VAEs, diffusion models \cite{sohl2015deep, ho2020denoising, dhariwal2021diffusion} start to show success in creating images of quality even higher than GANs. Inspired by non-equilibrium thermodynamics, diffusion models define a Markov chain that gradually adds Gaussian noise to input and learns to reverse it for data construction, resulting in a relatively low sampling speed.
While GANs and diffusion models have shown promising results in image generation, VAEs' latent space manipulation enables efficient aggregation of group features without iterative optimization or computationally expensive dimension reduction. Recently, latent diffusion models \cite{rombach2022high} have emerged that combine the latent space into diffusion models, but the method is limited by its low sampling for real-time template generation. Hence, VAE is uniquely suited for generating templates of scalable groups for its efficiency and interpretability.

\section{Methods}
In this section, we will introduce the structure of the proposed \textbf{InstantGroup}, a scalable and efficient groupwise template generation method. We present a comparison in Figure \ref{fig:overview} of previous learning-based methods (a, b) \cite{ding2022aladdin} and the proposed method (c, d) to highlight the novelty of InstantGroup. (a) and (b) both construct frameworks to update the parameters of the template by minimizing the loss function, which requires training on the test group for an optimal template, while InstantGroup provides different procedures for the training (c) and testing stage (d). In the training stage (c), a pair of scans is fed to the encoder-decoder simultaneously, and the networks are updated by losses involving the inputs and a temporary template image. In the testing stage (d), the number of inputs (the group size) is not fixed or limited; all group subjects will be encoded, and the template is then decoded from the average latent representation.
The detailed architecture of InstantGroup is displayed in Figure \ref{fig:framework} with Dual VAE, the Displacement Inversion Module (DIM), and the Subject-Template Alignment Module (STAM). The Dual VAE is the backbone with dual-path VAEs of shared weights to process pairs of input simultaneously. Then the output of the Dual VAE and the pair of inputs will be fed to DIM and STAM forwardly. These modules will be introduced subsequently in this section.

\subsection{Dual VAE Backbone}
% Dual-VAE 
The proposed method utilizes the widely used generative model named variational autoencoders (VAEs) \cite{kingma2013auto, rezende2014stochastic} to encode the brain MRI scans into a latent representation and decode them back. 
The encoder, parameterized by $\Phi$, encodes input $x$ into an approximated multivariate latent distribution $q_\Phi(z|x)$. The decoder, parameterized by $\theta$, reconstructs the image $\hat{x} \sim p_\theta(x|z)$ from the sampled latent representation vector $z \sim q_\Phi(z|x)$. Both the encoder and decoder employ 3D convolutional neural networks.
The optimization objective is to maximize the reconstruction likelihood between $x$ and $\hat{x}$ while minimizing the KL divergence between the learned posterior $q_\Phi(z|x)$ and the assumed normal prior distribution $p(z)$. A scaling factor $\beta$ \cite{higgins2017beta} is applied to the KL term for more flexible latent space representation:
\begin{equation}	
\mathcal{L}(x;\Phi, \theta) = E_{q_\Phi(z|x)}[log p_\theta(x|z)] - \beta KL(q_\Phi(z|x)||p_\theta(z)),
\end{equation}

As shown in Figure \ref{fig:overview} (d), we leverage the arithmetic on latent vectors of VAEs to generate the explicit common center for MRI groups of arbitrary size, which overcomes the fixed-size limitation of previous learning-based methods. Given the group of $N$ MRI subjects $I_1, I_2, ..., I_N$ over space $\Omega \subset \mathbb{R}^3$ in the testing stage, we first feed them separately (can be in parallel) into the encoder and obtain their latent vectors $z_1, z_2, ..., z_N$. Second, compute the average vector $\bar{z} = \frac{1}{N} \sum_i^N z_i$, which is the centroid of the representatives. Finally, the decoder constructs the template from $\bar{z}$ for GIR.

Based on VAEs, we propose \textbf{Dual VAE} that concurrently train two sets of VAEs that share the same networks and weights with a pair of input scans $I_1$ and $I_2$ as shown in Figure \ref{fig:framework}. The encoder predicts the corresponding latent vectors $z_1$ and $z_2$ of the inputs. Then the decoder reconstructs the input scans together with their representative template $I_{temp}$ from $\bar{z}$. The loss term of the Dual VAE is as follows: 
\begin{equation}
    \mathcal{L}_{DVAE} = \lambda_{KL}\sum_{i=1}^2 \mathcal{L}_{KL_i} + \lambda_{recon} \sum_{i=1}^2 \mathcal{L}_{recon_i},
\end{equation}
where $\mathcal{L}_{recon_i}$ is the mean squared error between $I_i$ and reconstructed $I_i$. 

The key advantage of the Dual VAE backbone is to pave the way toward arbitrary numbers of inputs at the testing stage by processing two inputs at the training stage, which narrows the leap from $1 \to N$ to $2 \to N$ when switching from training to testing. Training with pairwise processing as the minimal group to mimic the GIR scenarios also enables the combination of advanced modules (to be introduced in subsequent sections) to enhance the reliability and scalability of the framework.

\subsection{Displacement Inversion Module}
The Dual VAE backbone learns a latent distribution of the dataset, enabling interpolation in the latent space and compression of a group to the template. However, the image decoded from the average latent vector is not an anatomically meaningful template that can be smoothly morphed between group subjects \cite{berthelot2018understanding}. 
A template must be not only semantically representative but also anatomically centered and unbiased. We introduce the {Displacement Inversion Module (\textbf{DIM}) with an embedded registration network to ensure the generation of anatomically unbiased templates, through which we encourage the interpolated outputs to be closer to brain MRI scans and correlate the latent interpolation with registration distances.

\begin{figure}
    \centering
    \includegraphics[width=1.\linewidth]{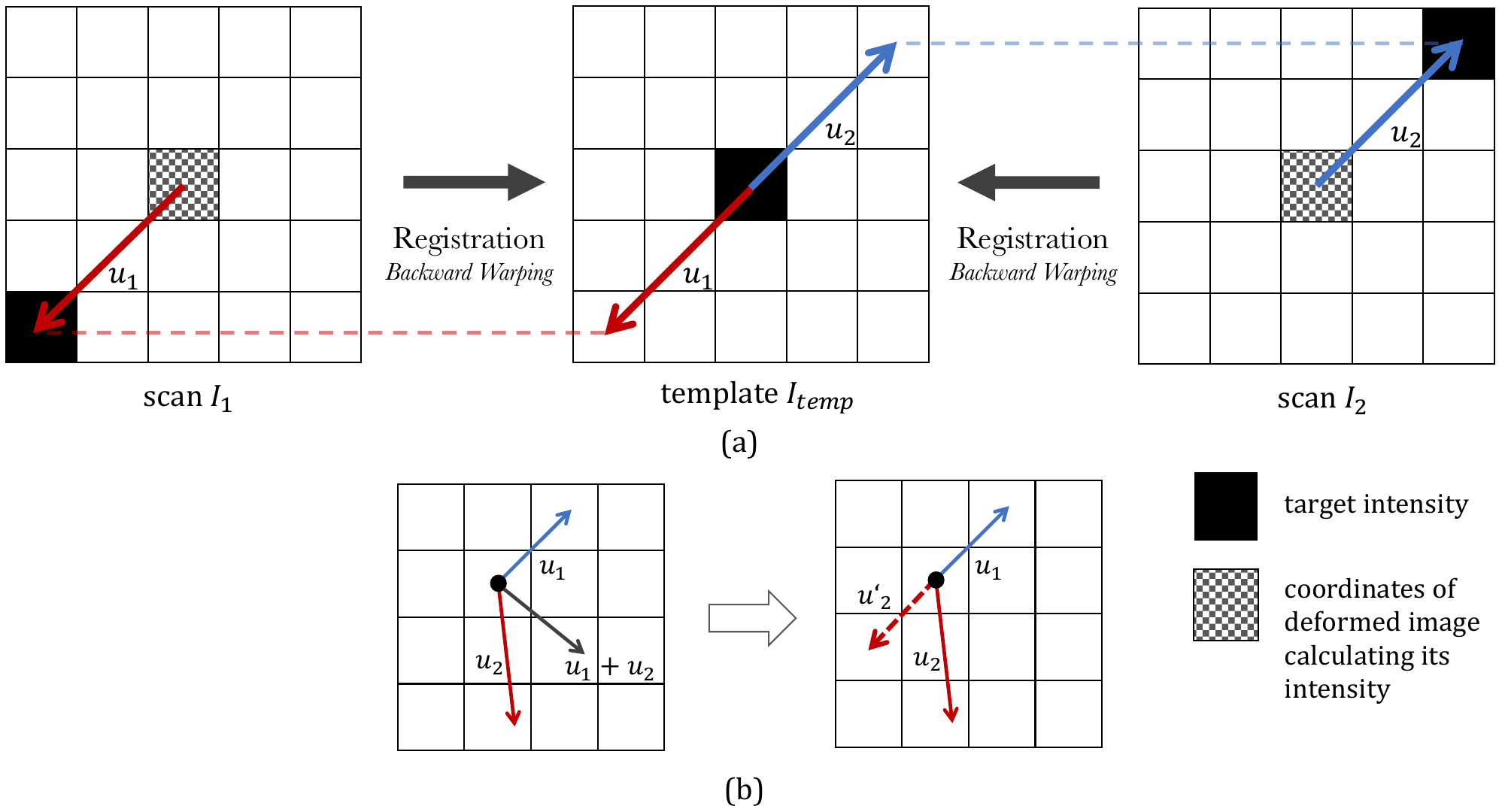}
    \vspace{-0.3cm}
    \caption{(a) Backward warping that determines the original location of target voxels to calculate the intensity through interpolation. When registering $I_1$ to $I_{temp}$, the displacement field indicates that the center pixel should be sampled from the value around the lower left corner, which is the inverse of that of $I_2$ to $I_{temp}$. (b) To minimize $||u_1 + u_2||^2$, the optimal vector is $u_2'$ instead of $u_2$. Here, we use the notation $u$ to denote the displacement vector at certain coordinates.}
    \label{fig:loppo}
    \vspace{-0.2cm}
\end{figure}

For two inputs $I_1$ and $I_2$, InstantGroup generates the template image $I_{temp}$ from the average latent vector. We perform pairwise registration from $I_1$ to $I_{temp}$ and $I_2$ to $I_{temp}$ using stationary velocity field (SVF) based registration methods \cite{ashburner2007fast, hernandez2009registration, shen2019networks, dalca2018unsupervised, dalca2019learning} that model the deformation field $\phi$ by integrating the predicted stationary velocity field $v$ over time $t=[0, 1]$. The ordinary differential equation is given as follows:
\vspace{-0.1cm}
\begin{equation}
\frac{\partial \phi_v^t}{\partial t} = v\circ \phi_v^t, 
\end{equation}
\vspace{-0.1cm}
where $\phi^0$ is the identity map. The neural network $Reg$ takes two input images as the moving $I_m$ and fixed $I_f$ and predicts the stationary velocity field $v = Reg(I_m, I_f)$. Using \emph{scaling and squaring} \cite{arsigny2006log, krebs2018unsupervised, dalca2019unsupervised}, we can numerically calculate the integration $\phi^1$ (or $\phi$) that deforms $I_m$ to $I_f$. Diffeomorphism has the property of invertibility. The inverse deformation field $\phi^{-1}$ that deforms $I_f$ to $I_m$ can be computed by integrating the negative velocity field $\phi_v^{-1} = \phi_{-v}$. The differentiable spatial transform module $STN$ computes each (sub)voxel linearly interpolated from neighboring voxels at the mapped location through the deformation field $\phi$. Let $u$ be the displacement field representing the vector difference between the corresponding points in the moving and fixed images with $\phi = Id + u$,
where $Id$ is the identity transformation representing the original coordinates.

We adopt the displacement field to represent the geometric distance or variant between the template and input scans. As illustrated in Figure \ref{fig:loppo}, backward image warping solves the coincidence by assigning the original coordinates for the target pixels to refer to. We propose that the displacement vectors from all group subjects at each voxel of $I_{temp}$ shall cancel out, resulting in the template voxel in equilibrium, which indicates its center position with even deformation distances to each subject, focusing on local discrepancies at the voxel level. 

Combined with the Dual VAE backbone, to enforce the two displacement fields from $I_1$ and $I_2$ to the template to be inverse, i.e., of the same magnitude in opposite directions, we propose in \textbf{DIM} the loss term of the squared L2-norm of sum of $u_1$ and $u_2$ to enforce an similar distance between $I_{temp}$ and each input, 
\begin{equation}
	\mathcal{L}_{DIM} = ||u_1 + u_2||^2.
\end{equation}
Through \textbf{DIM} and $\mathcal{L}_{DIM}$, the framework explicitly aligns the geometry of the latent space with the displacement fields. The template, decoded from the average of two latent vectors, is supposed to be equidistant from the two inputs in the deformation field space rather than being positioned at an implicit, unknown location.

\subsection{Subject-Template Alignment Module}
% Inverse Alignment 
The Dual VAE backbone generates a pair of subjects and the corresponding template, constituting the minimal unit of a group. 
To leverage this feature for improving the quality and reliability of the generated template, we introduce the Subject-Template Alignment Module (\textbf{STAM}). 
This module ensures the reliable generation of the templates by maximizing the similarity between the deformed templates and their corresponding moving inputs.

The three potential comparison combinations are \textbf{(1)} similarity between the deformed template and each input; \textbf{(2)} similarity between the deformed input scan and the template; \textbf{(3)} similarity between the deformed scans (towards the template).
Considering the nature of generated templates as a form of the average representative is its less clear appearance, comparing deformed scans in the template space as (2) or (3) may not be accurate enough during the registration process. In addition, denoting the energy functional of minimizing the mean squared error between the deformed template and the input scan as $E$, using Gâteaux Variation for the template $I_{temp}$, it can be derived \cite{ding2022aladdin}:
\vspace{-0.1cm}
\begin{multline}
    \delta E(I_{temp}; dI_{temp})  = 2 \langle \sum_{i=1}^{N} (I_{temp} - I_i \circ \phi_i) |D\phi_i|, dI_{temp} \rangle \\
 \overset{!}{=} 0, \, \forall dI_{temp},
\end{multline}
\vspace{-0.1cm}
with the optimal template $I_{temp}^*$ being:
\begin{equation}
I_{temp}^* = \frac{\sum_{i=1}^N I_i \circ \phi_i |D\phi_i|}{\sum_{i=1}^N |D\phi_i|},
\end{equation}
where $D$ denotes the Jacobian. The optimal template indicated constraints on the similarity between inputs, and the deformed template would push the template to a weighted average depending on the deformation fields. Therefore, we apply a constraint $\mathcal{L}_{STAM}$ that measures the similarity between the deformed template $I_{temp} \circ \phi_i$ and each scan in the input space:
\begin{equation}
	\mathcal{L}_{STAM} = \sum_{i \in \{1,2\}} \mathcal{L}_{similarity}(I_{temp} \circ \phi_i, I_i),
\end{equation}
where $\phi_k$ represents the deformation field from $I_{temp}$ to $I_k$, and $\mathcal{L}_{sim}$ notes the dissimilarity measurement, which is the mean squared error. 

$\mathcal{L}_{STAM}$ ensures the transitions between $I_{temp}$ and $I_1$, $I_2$ and therefore enhances the continuity of the latent space, as the loss function constraints that small changes in the latent vector lead to a new vector that still lies on the manifold and is able to decode to an authentic image.

\subsection{Training and testing settings, and implementation}
\subsubsection{Training stage}
The framework is trained with Dual VAE as shown in Figure \ref{fig:framework} and optimized by the overall loss function. 
In the training stage, two scans are fed into the networks and processed through dual symmetric paths via the weight-sharing encoder, decoder, and registration module. The diffeomorphic registration network \textbf{Reg} is pre-trained and frozen during the training of InstantGroup. After the generation of the template $I_{temp}$ from $I_1$ and $I_2$, \textbf{Reg} predicts the deformation fields from $I_{temp}$ to $I_1$ and $I_2$ correspondingly. \textbf{DIM} incorporates the deformation fields and forces the displacements to be of the same magnitude and opposite. \textbf{STAM} deforms $I_{temp}$ with the deformation fields and enforces the similarity between $I_1$, $I_2$ and the deformed $I_{temp}$. The framework is optimized by the overall objective loss function:
\begin{equation}
\label{loss}
\mathcal{L}(I_1, I_2) = \mathcal{L}_{DVAE} + \lambda_{dim} \mathcal{L}_{DIM} + \lambda_{stam} \mathcal{L}_{STAM}.
\end{equation}
\subsubsection{Testing stage}
The testing stage is displayed in Fig. \ref{fig:overview} (d). Please note that at the testing stage for template generation, InstantGroup does not need to utilize \textbf{Reg}, \textbf{DIM}, and \textbf{STAM}. Therefore, the framework breaks the limitation of two inputs during the training stage and can accept an arbitrary number of inputs for encoding. Given a group of $N$ MRI subjects, we encode them to latent vectors $z_1, ..., z_N$ and average $\bar{z} = \frac{1}{N}\sum z_i$. The decoder reconstructs the template image from $\bar{z}$. Therefore, InstantGroup can be applied to groups of different sizes.
\subsubsection{Implementation}
We implement the proposed InstantGroup using PyTorch \cite{paszke2017automatic}.  
The encoder is comprised of four 3D convolution modules with 128, 128, 128, and 256 filters. Each module contains a 3D convolution (kernel size=3, stride =2, padding =1) followed by LeakyReLU. 
The encoder outputs the mean $\mu$ and log variance $\log(\sigma)$ of a normal distribution; the reparameterization trick \cite{kingma2013auto} is applied to sample vectors from the distribution. The latent representation $z$ is of shape $128\times 6 \times 7\times 6$. The decoder employs four 3D transposed convolutions (kernel size=3, stride =2, padding=1) with 128, 128, 128, and 128 filters with LeakyReLU activation. The final convolution is activated by the sigmoid. After random searching, we obtain the optimal hyperparameters as $\lambda_{recon} = 300, \lambda_{KL} = 0.0002, \lambda_{dim} = 7.5, \lambda_{stam} = 100$.  The training time for OASIS and ADNI is approximately 21 hours (6 hours for 300 epochs VAE pre-training and 15 hours for 100k iterations InstantGroup training) and 62 hours (14 hours for 300 epochs VAE training and 48 hours for 300k iterations InstantGroup training) respectively. Parameters are optimized using Adam with an initial 1e-4 learning rate and decreased by a cosine annealing schedule (T = 4). 
The diffeomorphic registration network is developed from the official implementation of VoxelMorph \cite{balakrishnan2019voxelmorph} with default hyperparameters and trained for 30k iterations.

\renewcommand{\arraystretch}{1.25}
\begin{table*}[ht]
\centering
\caption{Test time table (default unit is seconds if not specified). The runtime of Aladdin and DeformT refers to the time of fine-tuning on the test set. Values in the format of "\textgreater $x$" refer to a runtime much larger than $x$.}
\begin{tabular}{>{\centering\arraybackslash}p{1.6cm}|>{\centering\arraybackslash}p{1.3cm}>{\centering\arraybackslash}p{1.3cm}>{\centering\arraybackslash}p{1.3cm}>{\centering\arraybackslash}p{1.3cm}>{\centering\arraybackslash}p{1.3cm}>{\centering\arraybackslash}p{1.3cm}>{\centering\arraybackslash}p{1.3cm}} 
\hline
Group size & 2 & 4 & 8 & 16 & 32 & 64 & 128 \\
\hline
\textbf{InstantGroup}  & \textbf{1.350}    & \textbf{2.486}  & \textbf{5.128}  & \textbf{9.958}  & \textbf{20.55}  & \textbf{40.88}  & \textbf{81.56}  \\
\hline
ANTs    & 64.79   & 118.9  & 163.7    & 322.9    & 697.8    & 1223   & \textgreater 50 min   \\
\hline
ABSORB    & 14.74   & 55.99  & 225.4  & 505.7  & 1448   & \textgreater 50 min  & \textgreater 50 min  \\
\hline
Aladdin        & 97.13   & 195.5  & 394.7   & 797.8   & 1625 & 3358  & \textgreater 60 min\\
\hline
DeformT          & 442.9  & 960.1  & 1181  &  1506    & 2504     & 3987 & \textgreater 60 min \\
\hline
\end{tabular}
\vspace{-0.2cm}
\label{tab:runtime}
\end{table*}

\section{Experiments}
\subsection{Datasets}
%The proposed framework InstantGroup is evaluated on two public brain MRI scan datasets.
\subsubsection{OASIS}
The Open Access Series of Imaging Studies (OASIS-1: Cross-sectional MRI Data in Young, Middle Aged, Nondemented and Demented Older Adults) \cite{marcus2007open} consists of T1-weighted MRI scans obtained from 416 subjects aged from 18 to 96 with 48 annotated segmentation labels. All volumes were preprocessed \cite{yaniv2018simpleitk}, including skull stripping, intensity normalization, affine registration to MNI brain space (Colin27), cropped and resampled to $96\times112\times96$. 
The dataset is divided into 305/40/80 as training, validation, and test sets. 

\subsubsection{ADNI}
The Alzheimer's Disease Neuroimaging Initiative (ADNI) \cite{mueller2005ways} is a public database for research on Alzheimer's disease (AD), like detecting and tracking disease progression. We apply Sequence Adaptive Multimodal SEGmentation (SAMSEG) of FreeSurfer to perform automatic segmentation, followed by preprocessing similar to OASIS. The dataset is divided into 635/40/200 as training/validation/test sets.
For both datasets, the train/val/test sets are split by patients/subjects to avoid possible data leakage.

\subsection{Baseline methods}
We compare InstantGroup with two traditional baseline methods ABSORB \cite{jia2010absorb}, ANTs \cite{avants2010optimal}. ABSORB \cite{jia2010absorb} uses affinity propagation \cite{frey2007clustering} for clustering and bundling subjects using diffeomorphic demons \cite{vercauteren2009diffeomorphic} as the registration method. ANTs \cite{avants2010optimal} implemented in the ANTs software package \cite{avants2008symmetric}, estimates the optimal templates and optimizes the geometric component iteratively. The number of iterations is 4, and the gradient is 0.2. 

We also include two recent learning-based methods, DeformT \cite{dalca2019learning}, and Aladdin \cite{ding2022aladdin}. The framework of the two learning-based methods is also displayed in Figure \ref{fig:overview} (a) and (b). DeformT \cite{dalca2019learning} is a learning framework for building deformable conditional templates for a different group of subjects. 
We pre-train the network for 2000 epochs for around 14 hours; for each test group, we fine-tune the model for 20 $\sim$ 300 epochs until convergence;
Aladdin \cite{ding2022aladdin} performs jointly atlas building and registration with pairwise image alignment. We pre-train the network for 36 hours and fine-tune the model for around 40 $\sim$ 200 epochs until convergence for each test set.

Both DeformT\cite{dalca2019learning} and Aladdin\cite{ding2022aladdin} were originally designed to construct generic templates from large-scale datasets. In this study, we adapt them to generate group-tailored templates by fine-tuning the models on each test group individually for a more consistent and fair comparison with the proposed method InstantGroup.

\subsection{Evaluation metrics}
\subsubsection{Unbiasedness}
 To measure the unbiasedness degree of generated templates, we apply two metrics proposed in [4] that utilize the magnitude related to the deformation field (warping each group subject to the template) to evaluate this quantitatively: \textbf{Centrality} that computes the 2-norm of the mean displacement field and \textbf{AvgDisp}, which refers to the average displacement magnitude:
 \begin{equation}
 	Centrality = \frac{1}{N}||\sum_i^N u_i||_2, AvgDisp = 	\frac{1}{N}\sum_i^N ||u_i||_2.
 \end{equation}
For both metrics, lower values indicate higher unbiasedness.
 
\subsubsection{Smoothness}
To measure the degree of smoothness and to which extent the deformation field preserves topology, we use two metrics \textbf{$|$LogJ$|_{95}$} and \textbf{SDLogJ} related to the determinant of the Jacobian matrix $J_\phi(p) = |\nabla \phi(p)|$ that measures the local distortion around each voxel $p$. 
\textbf{$|$LogJ$|_{95}$} computes the 95\% percentile of the absolute value of the logarithm of the Jacobian determinant. \textbf{SDLogJ} computes the standard deviation of the logarithm of the Jacobian determinant. For both metrics, lower values indicate a smooth and plausible deformation field. We also evaluate the number of foldings by counting the negative Jacobian determinant $\{p: J_\phi(p) < 0\}$ to represent extreme local irregularities of the deformation field.

\begin{figure}
    \centering
    \includegraphics[width=0.6\linewidth]{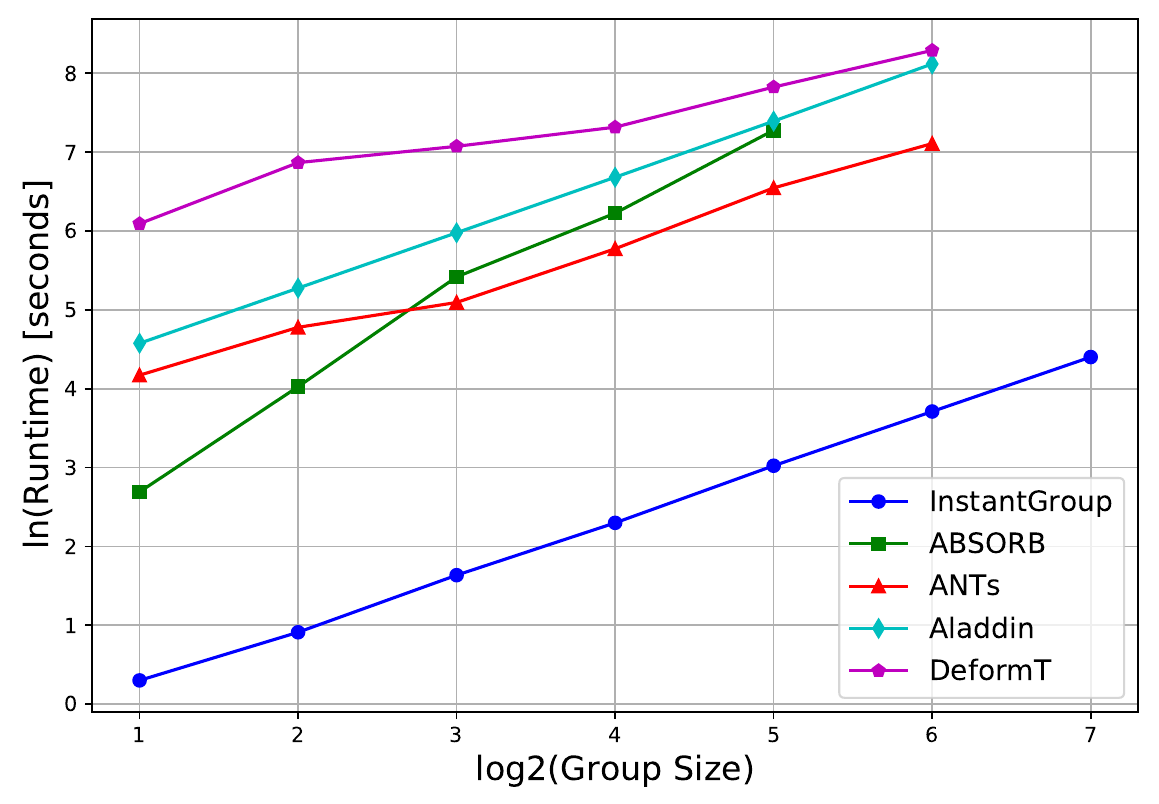}
    \vspace{-0.2cm}
    \caption{Runtime (seconds in natural logarithm) comparison of the proposed method and baselines across different group sizes (in the log with base 2). Note that runtime in Table \ref{tab:runtime} without exact values is omitted.}
    \vspace{-0.1cm}
    \label{fig:testtime}
\end{figure}

\subsubsection{Dice score and Hausdorff distance}
In addition to Centrality and AvgDisp evaluating the unbiasedness of the generated templates, we also evaluate the registration accuracy with the template. 
We apply the widely used Dice Similarity Coefficient \textbf{DSC} to evaluate the template quality and groupwise registration performance, indicating the degree to which subjects are warped to the same space. To eliminate the impact of potential template quality deviations, we compute the average anatomical label overlap between each warped group subject pair instead of that between subjects and templates. The formula is given as follows:
\vspace{-0.1cm}
\begin{equation}
	DSC = \frac{1}{N^2 N_s}\sum_{i, j}\sum_{k}^{N_s} \frac{2\times|W_i^k \cap W_j^k|}{|W_i^k| + |W_j^k|},
\end{equation}
\vspace{-0.1cm}
where $N$ refers to the group size, $N_s=17$ refers to the number of labels, $W_i^k$ is the pixel set of $k_{th}$ region of $i_{th}$ warped subject. Higher scores refer to better performance. 
The labels utilized are the 17 largest anatomical structures of the brain, including: Cerebral Cortex, Cerebral White Matter, Cerebellum Cortex, Lateral Ventricle,  Cerebellum White Matter, Brain Stem, Thalamus, Putamen, Hippocampus. All structures comprise left and right regions except the Brain Stem.

We also utilize the 95\% percentile of the Hausdorff distance \textbf{HD95} to measure the distance between each pair of warped group subjects to reflect the robustness of the GIR method.

\renewcommand{\arraystretch}{1.4}
\begin{table*}[ht]
\fontsize{7.5pt}{7.5pt}\
\centering
\caption{Quantitative results of the experiments on OASIS and ADNI with six metrics. $\uparrow$ denotes better performance with higher values and $\downarrow$ denotes better performance with lower values. For the mean determinant of the Jacobian matrix, values closer to 1 indicate better results.
"Affine" refers to subjects before undergoing GIR. Results in bold indicate the best performance for each metric (significantly better than the second best with $p<0.05$ in paired T-Test). 
}
\label{table-results}
\setlength{\tabcolsep}{3.5pt}
\begin{tabular}{l|llllll|llllll} 
%\small
\toprule
\multirow{2}{*}{Methods} & \multicolumn{6}{c|}{OASIS} & \multicolumn{6}{c}{ADNI}  \\ \cline{2-13}
 & Central $\downarrow$ & AvgDisp $\downarrow$ & DSC $\uparrow$ & HD$_{95}\downarrow$ & $|$LogJ$|_{95}$ & SDLogJ &  Central $\downarrow$ & AvgDisp $\downarrow$ & DSC $\uparrow$ & HD$_{95}\downarrow$ &  $|$LogJ$|_{95}$ & SDLogJ \\ 
\hline
Affine &  - & - & 60.12 $\pm$ 1.45 & 2.308 $\pm$ 0.09  & - & - 
&- & - & 62.96 $\pm$ 2.56 &2.144 $\pm$ 0.18  & -  & - \\
\hline
ANTs & 239  $\pm$ 10    & \textbf{873.7} $\pm$ 37  & 73.55 $\pm$ 0.44 & 1.478 $\pm$ 0.04  & 0.0015 & 0.1349 
& 348 $\pm$ 19  & 936.8 $\pm$ 28 & 74.95 $\pm$ 3.09 & 1.282 $\pm$ 0.18 & 0.0016 & 0.1351        \\

ABSORB & 743 $\pm$ 57 & 1102 $\pm$ 88 & 73.50 $\pm$ 0.55 & 1.450 $\pm$ 0.04   & 0.0021    & 0.1697
& 870 $\pm$   76    & 1198 $\pm$ 94 & 74.87 $\pm$ 2.92  & 1.276 $\pm$ 0.16   & 0.0024 & 0.1836  \\

\hline
DeformT & 353 $\pm$ 36  & 943.4  $\pm$ 35  & 74.02 $\pm$ 0.35 & 1.427 $\pm$ 0.03 & 0.0016 & 0.1546
 & 359 $\pm$ 51 & 946.1 $\pm$ 37 & 74.98 $\pm$ 3.45 & 1.280 $\pm$ 0.17 & 0.0016 & 0.1597 \\ % 0.004 & 0.019
 
Aladdin& 585 $\pm$ 26 & 1035  $\pm$  53 & 74.04 $\pm$ 0.42 & 1.431 $\pm$ 0.03  & 0.0018 & 0.1846
& 682 $\pm$ 45 & 1099 $\pm$ 39 & 74.25 $\pm$ 2.76 & 1.297 $\pm$ 0.14  & 0.0018   & 0.1819\\ % 0.026  % 0.005 
\hline
\textbf{InstantGrp}& \textbf{168} $\pm$ 3.0  & 895.6 $\pm$ 36     & \textbf{74.16} $\pm$ 0.38 & \textbf{1.414} $\pm$ 0.03  & 0.0015 & 0.1394
& \textbf{179} $\pm$ 17 & \textbf{920.2} $\pm$ 25 & \textbf{75.31} $\pm$ 3.10 & \textbf{1.267} $\pm$ 0.17 &  0.0017 & 0.1417  \\

\bottomrule
\end{tabular}
\end{table*}

\begin{figure*}[!t]
\centerline{\includegraphics[width=2\columnwidth]{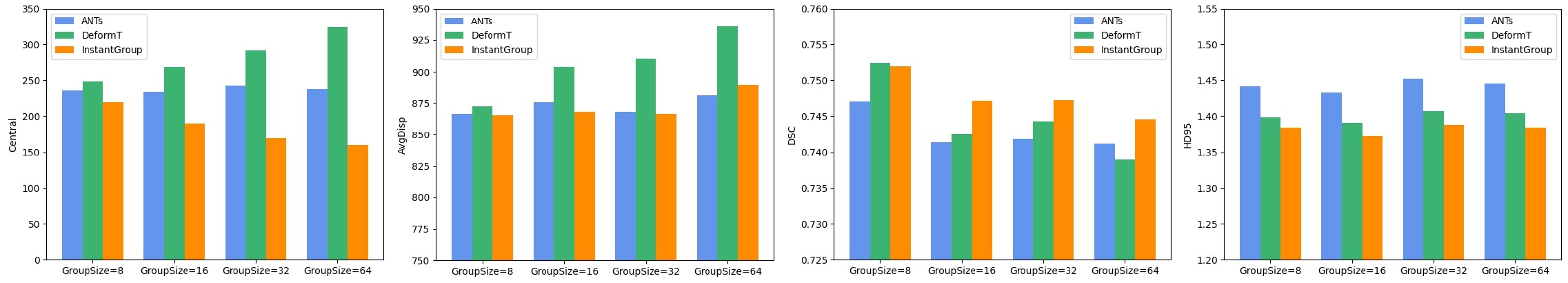}}
\vspace{-0.2cm}
\caption{Test groups of sizes 8, 16, 32, 64 compared to the two representative benchmarks ANTs and DeformT on the four metrics Centrality, AvgDisp, DSC, and HD95. InstantGroup has comparable or superior robust performance across all group sizes.}
\label{multigs}
\vspace{-0.1cm}
\end{figure*}

\subsection{Runtime Analysis}
We present the runtime of different group sizes ranging from 2 to 128 in Table \ref{tab:runtime}. The proposed method, InstantGroup, exhibits significant runtime improvement over traditional and other learning-based baselines across all tested group sizes. 

For small groups like 4 or 8 subjects, InstantGroup generates the template in just seconds. As the group size increases, InstantGroup continues to scale efficiently. The runtime grows almost linearly with the group size, showcasing its scalability.
For instance, InstantGroup can process a group of 128 subjects in just around 81 seconds, while ABSORB and ANTs take dramatically longer over 3000 seconds and DeformT exceeding 60 minutes to converge. 
Compared with traditional methods, InstantGroup can process 128 subjects in less time than ABSORB or ANTs takes to process just 8 subjects. 

For the learning-based methods, the runtime is counter-intuitively longer. This inefficiency arises as Aladdin and DeformT involve training or fine-tuning the parameters to optimize the templates until convergence, which demands substantial runtime. 
When the group size increases to 128, Aladdin and DeformT take over 60 minutes to process such large groups, making InstantGroup over 40 times faster.

InstantGroup presents a significant reduction in runtime compared to other methods, which often take several minutes to hours to achieve the same task. The linear scalability of InstantGroup is a direct result of its design. As the group size increases, the performance gap between InstantGroup and other methods becomes even more pronounced. 

Figure \ref{fig:testtime} plots the logarithm of the runtime comparison. InstantGroup stands out as the most efficient method able to scale linearly with group size. Such significant runtime savings make InstantGroup highly suitable for practical applications in medical image analysis, where time efficiency is crucial.

\subsection{Computational Costs}
All learning-based methods are trained and tested on a single Nvidia 2080Ti GPU. The number of trainable parameters of InstantGroup, Aladdin, and DeformT are 1.3, 2.0, and 1.1 million. 
The GPU memory usage of them is 11, 2.5, and 2.9 GB. The memory usage of InstantGroup is greater than that of the learning-based baseline methods, but it's still affordable for professional use.

\section{Quantitative results}
Table \ref{table-results} presents a comprehensive quantitative comparison of InstantGroup with baseline methods on the two datasets, OASIS and ADNI. The results indicate that InstantGroup consistently outperforms the baselines in terms of Centrality (Central), Average Displacement (AvgDisp), Dice Similarity Coefficient (DSC), Hausdorff Distance (HD95), $|$LogJ$|_{95}$, and SDLogJ.
\subsection{Performance on unbiasedness and smoothness}
For Centrality and AvgDisp, which assess the degree of unbiasedness and geometric center alignment in terms of the deformation fields, InstantGroup demonstrates superior performance on both datasets. Specifically, InstantGroup achieves a Centrality value of 168 on OASIS and 179 on ADNI (29.7\% and 48.6\% improvement), significantly lower than all other methods. This indicates that the displacement fields from each group subject to the template are more in equilibrium, and templates generated by InstantGroup are more evenly centered within the group with minimal bias toward any particular subject. Additionally, the AvgDisp of InstantGroup on OASIS is 895.6, 2.3\% higher than the best, and the AvgDisp of InstantGroup on ADNI is 920.2, 1.8\% lower than the second best. 
The quantitatively results validates the robustness of InstantGroup in generating templates that minimize the magnitude of displacements, leading to uniformity and consistency in the GIR process.

As we measure the smoothness of the deformation fields, no methods have negative Jacobian determinants, meaning all of the deformation fields are folding-free. 
$|$LogJ$|_{95}$ and SDLogJ are presented in Table \ref{table-results}. InstantGroup achieves low values for both metrics, indicating satisfying smoothness of the deformation fields. The results remain comparable to other baselines while slightly higher than the values of ANTs, which could be partially considered a trade-off for its improved accuracy in alignment by InstantGroup. This indicates that the templates generated by InstantGroup lead to a balance between topology-preserving deformations and registration performance.

\begin{figure}[!t]
\centerline{\includegraphics[width=0.9\columnwidth]{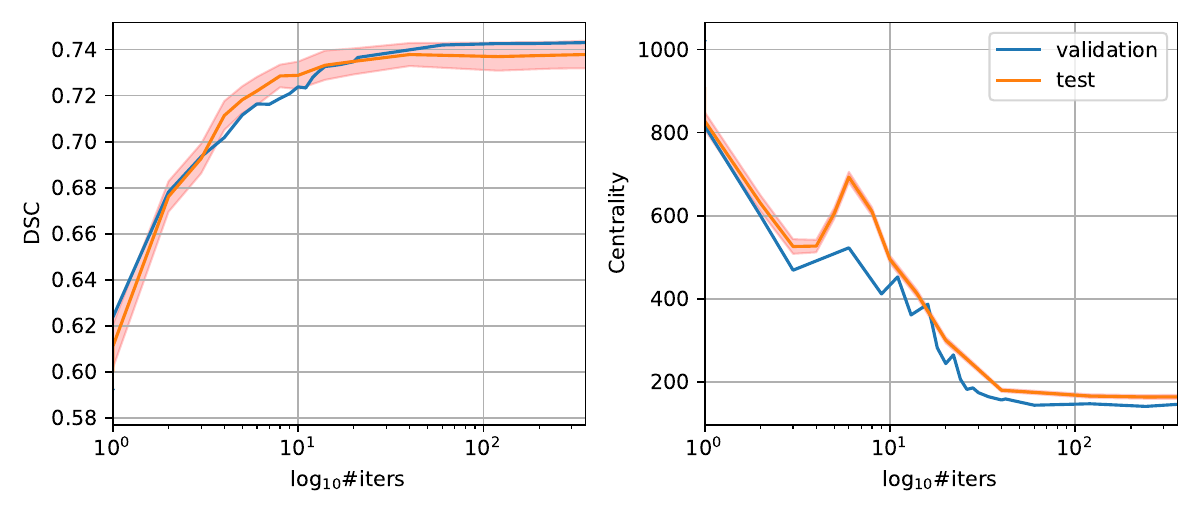}}
\vspace{-0.2cm}
\caption{The curve of GIR performance on the validation group and test groups during the training stage. The results of the test set are presented with the standard deviation band. \textbf{\#iters} refers to $10^3$ iterations.}
\vspace{-0.2cm}
\label{fig_learning_process}
\end{figure}

\begin{figure*}[!b]
\centerline{\includegraphics[width=1.5\columnwidth]{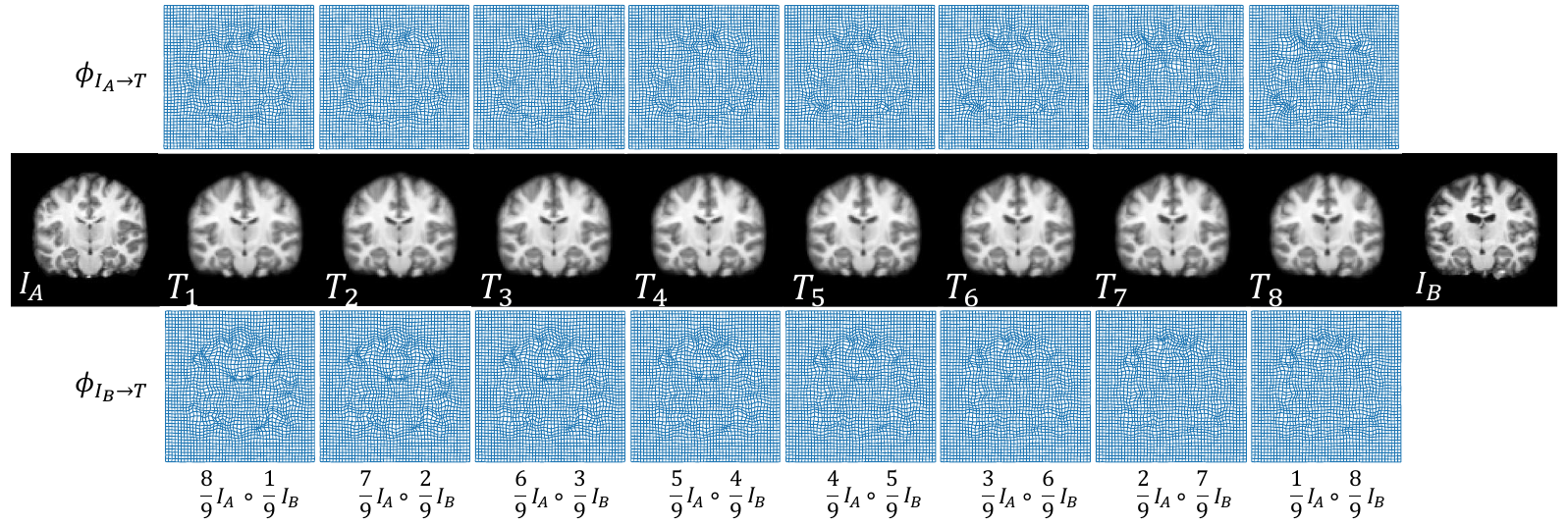}}
\vspace{-0.2cm}
\caption{Interpolation between two pairs of MRI scans $I_A$ and $I_B$. The interpolated images $T_k$ displayed in row 2, column 2 to 9, are decoded from $\frac{k}{9}z_A + \frac{9-k}{9}z_B$, where $z_A, z_B$ refer to the latent representation of the input $I_A$ and $I_B$. Rows 1 and 3 present the deformed grids of the displacement fields from $I_A$ to $T_k$ and from $I_B$ to $T_k$. The deformed grids are down-sampled by two for better visual effects.}
\vspace{-0.1cm}
\label{interpolation}
\end{figure*}

\subsection{Performance on anatomical alignment}
InstantGroup achieves the highest DSC and the lowest HD95 on both datasets, indicating superior registration performance. Specifically, InstantGroup improves the DSC by approximately $0.16\%$ over the second-best method (Aladdin) and reduces the HD95 by 0.91\% compared to DeformT on the OASIS dataset. On the ADNI dataset, InstantGroup improves the DSC by 0.48\% compared to DeformT and HD95 by 0.71\% compared to ABSORB. All of the improvements are statistically significant, with p-values smaller than 0.05 in a paired t-test. 

It is worth noting that the relatively modest improvements in DSC and HD95 are due to the fact that all methods use the same pairwise registration model to register group subjects to their templates. Since DSC and HD95 are more heavily influenced by the registration algorithm than the template generation process (which is the primary task of InstantGroup and baseline methods), the difference between InstantGroup and baselines can be minor. In addition, groupwise DSC tends to be lower compared to pairwise DSC since the subjects are registered to the medium (template) rather than being registered directly to each other. This characteristic inherently results in lower values and minor differences in DSC and HD95 among these methods in the GIR evaluation. 

\subsubsection{Multiple group size performance analysis}
Figure \ref{multigs} shows the scalability of InstantGroup and two representative benchmarks, ANTs and DeformT, across various group sizes (8, 16, 32, and 64).
InstantGroup maintains remarkable robustness and consistently demonstrates superior or comparable performance across all group sizes with higher DSC values and lower HD95 values. More critically, InstantGroup achieves low Centrality and AvgDisp across all group sizes indicating the template is a consistent center reference for each group. The improvement in Centrality is larger as the group size doubles to 32 and 64 when ANTs and DeformT struggle with larger groups. The comparison underscores the advantages of InstantGroup in consistently generating unbiased templates with competitive registration accuracy.

\subsubsection{Training stage curve}
We present the performance curve of InstantGroup on the validation and test groups in Figure \ref{fig_learning_process} during training. The figure indicates that InstantGroup converges effectively. 
The model exhibits significant improvement during the early stage, followed by meticulous optimization through further training and stabilizing at a high value. The Centrality oscillates during the early stage, which could be attributed to the notable enhancement in initial template quality and pairwise registration performance. The Centrality is then decreased steadily. By visualizing the performance curve throughout the training, we provide insights into the stable optimization of the proposed method.

\begin{figure*}[!htb]
    \centering
    \hspace*{-0.85cm}\includegraphics[width=0.95\linewidth]{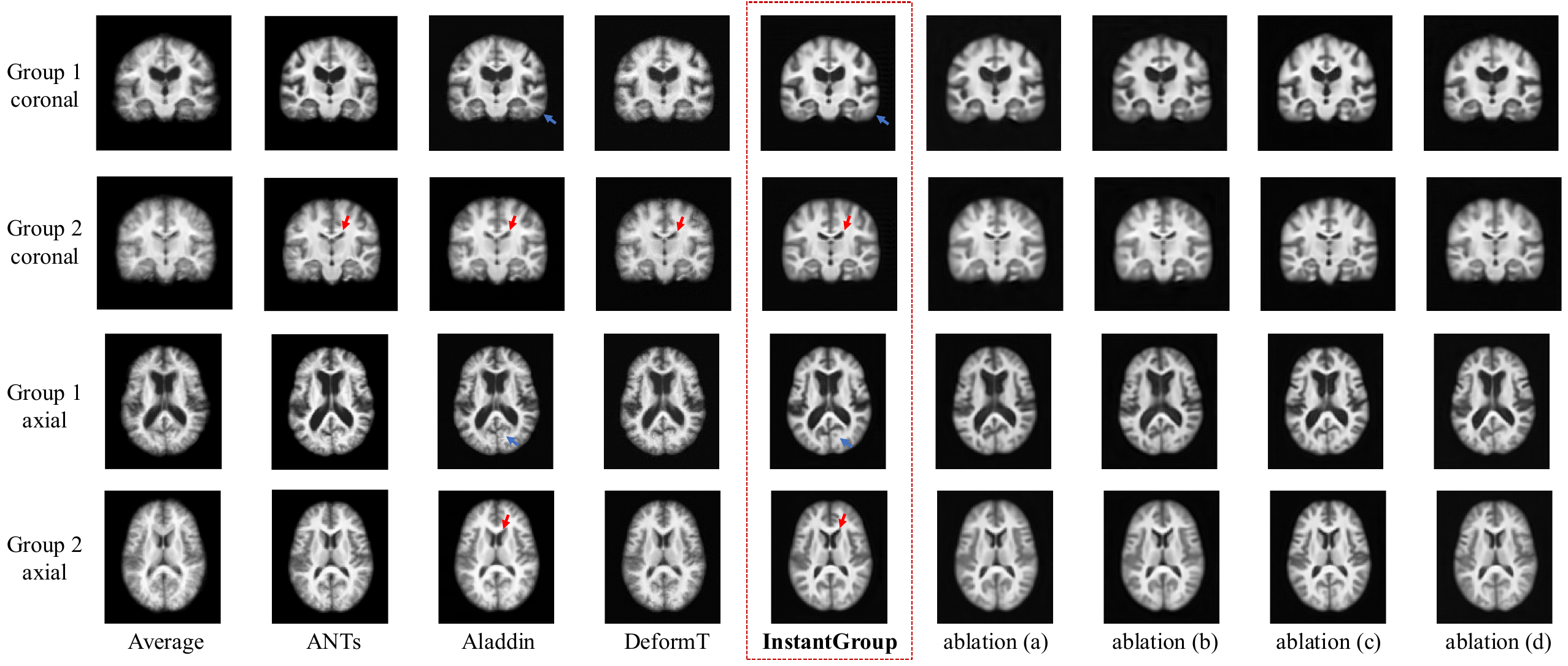}
    \vspace{-0.2cm}
    \caption{Qualitative results of baseline methods, InstantGroup, and ablation studies. Images displayed from left to right are: Average refers to the average of the group scans, templates by ANTs, Aladdin, DeformT, the proposed InstantGroup, and ablation (a) - (d) refer to the variants of InstantGroup for the ablation study. Two groups are presented. Group 1 is a series of scans with relatively large lateral ventricles, and group 2 represents scans with relatively small lateral ventricles. We annotated regions with red and blue arrows to indicate relatively good generation and poor generation of InstantGroup.}
    \vspace{-0.1cm}
    \label{fig:quality}
\end{figure*}

\subsection{Interpolation analysis and qualitative results}
Figure \ref{interpolation} demonstrates that InstantGroup can interpolate the latent vectors of input images smoothly and generate semantically meaningful fusions. The interpolation process between a pair of MRI scans ($I_A$ and $I_B$) illustrates how the method preserves anatomical continuity and structural details while transitioning from one scan to another. The magnitude of displacement fields $\phi_{I_A \to T}$, $\phi_{I_B \to T}$ from one input to the interpolation images ($T_1$ to $T_8$) increases proportionally with the distance in the latent space as the interpolation progresses.

Evaluating template quality is challenging since segmenting the template itself may not be accurate, since the template, being a form of the semantic average, often lacks clear anatomical structures \cite{ding2022aladdin}. We can visually compare the templates generated by baselines and the proposed method in Figure \ref{fig:quality}. We present two groups of four scans with different characteristics, group 1 with large lateral ventricle areas and group 2 with generally smaller lateral ventricles for comparison. For reference, the leftmost column exhibits the average of the group scans.
It could be observed that InstantGroup and all the baselines are able to generate high-quality images with clear anatomical structures compared with the average and reflect the group features well. The differences between InstantGroup and baselines are 1) InstantGroup can better delineate the edge of tiny but important regions, like when the lateral ventricles are small (indicated by red arrows in the figure), and 2) InstantGroup does not preserve as many details of gray matter area as baselines (indicated by blue arrows in the figure), this is due to inevitable information loss during the compression and the semantic average can cancel out the details of the scans. 
While the authenticity and details of InstantGroup templates may not be comparable to real MR scans, they preserve overall anatomical structures well, reducing high-frequency noise similar to the performance of baselines.
Moreover, sharper templates do not necessarily guarantee better GIR results. The primary objective of GIR is to warp subjects to an unbiased center instead of the sharpness of the template. 
Overall, InstantGroup exhibits a remarkable ability to generate semantically unbiased and anatomically consistent templates effectively.

\renewcommand{\arraystretch}{1.2}
\begin{table*}[!h]
\centering
\caption{Ablation study on group size 8 and 30 of each component of InstantGroup.(a) refers to a naive GIR method that iteratively takes the average of deformed subjects. (b-d) represent combinations of each module. (f-h) refer to adding an extra loss $\mathcal{L}_{pair}$.(e) is the proposed method InstantGroup.(i) replaces $\mathcal{L}_{DIM}$ by the other variant $\mathcal{L}_{DIM-abl}$.}
\label{table-ablation}
\setlength{\tabcolsep}{3pt}
\begin{tabular}{c|cccc|ccc|ccc}
\toprule
\multirow{2}{*}{Method} & \multirow{2}{*}{DualVAE} & \multirow{2}{*}{STAM} & \multirow{2}{*}{DIM} & \multirow{2}{*}{$\mathcal{L}_{pair}$} & \multicolumn{3}{c|}{group size = 8} & \multicolumn{3}{ c}{group size = 30} \\ \cline{6-11} 
  &  &  &  &  & Centrality$\downarrow$    & AvgDisp$\downarrow$    & DSC$\uparrow$    & Centrality$\downarrow$  & AvgDisp$\downarrow$  & DSC$\uparrow$  \\ 
\hline
a & & & & & 853.4 & 1214 & 75.60 & 588.7 & 1032 & 73.87 \\
b & \checkmark & & & & 338.8 & 871.3 & 75.05 & 352.0 & 946.2 & 73.86 \\
c & \checkmark & \checkmark & & & 299.3 & 874.3 & 75.47 & 343.8 & 965.8 & 73.97 \\
d & \checkmark & & \checkmark & & 198.7 & 809.4 & 75.51 & 182.5 & 894.8 & 73.89 \\
\hline
\textbf{e (InstantGroup)} & \checkmark & \checkmark & \checkmark & & \textbf{184.8} & \textbf{805.7} & \textbf{75.85} & \textbf{152.9} & 867.7 & \textbf{74.33} \\
\hline
f & \checkmark & & \checkmark & \checkmark & 215.2 & 834.9 & 75.49 & 206.9 & 866.9 & 73.53 \\
g & \checkmark & \checkmark & & \checkmark & 287.1 & 880.5 & 75.61 & 291.9 & 916.3 & 73.86 \\
h & \checkmark & \checkmark & \checkmark & \checkmark & 236.6 & 852.0 & 75.82 & 198.1 & 889.7 & 74.26 \\
\hline
i & \checkmark & \checkmark & $\mathcal{L}_{DIM-abl}$ & & 208.3 & 814.9 & 75.66 & 205.8 & \textbf{844.9} & 73.61 \\
\bottomrule
\end{tabular}
\label{tab2}
\end{table*}

\subsection{Analysis of network design}
\subsubsection{Ablation study on each component}
We conduct an ablation study on the validation group to investigate the impact of each module and loss term within InstantGroup, as shown in Table \ref{table-ablation}.
Method (a) refers to a naive baseline that iteratively generates the template by taking the average of all (deformed) group subjects.
Method (b) refers to utilizing the latent arithmetics with Dual VAE, which help ensure the continuity of the latent space and improve the unbiasedness of the generated templates with lower Centrality and avgDisp values. 
Adding STAM (c) does not have a significant improvement in the evaluation metrics aspect while improving the quality of the generated image with clearer contours and details. 
We can observe a significant reduction in Centrality and AvgDisp when introducing DIM (d) for its constraint on the displacements. However, the image quality degrades compared with (c). 
The method (e) InstantGroup that incorporates all components yields the highest DSC and overall lowest Centrality and AvgDisp values. The image quality of InstantGroup is close to (c) and significantly better and sharper than (a), (b), and (d); this indicates that good visual quality is mainly contributed by STAM, which enforces the similarity between the deformed templates and inputs.
The quantitative results indicate the successful integration of STAM and DIM for their mutual promotion: STAM ensures meaningful interpolation in the latent space to make DIM perform better and steadily.

The architecture difference between InstantGroup and (a) can be summarized as that the former aggregates the group subjects on the level of latent space, while the latter simply does so on the pixel level.
The difference results in huge gap in terms of the unbiasedness, proving InstantGroup's superiority ensuring the generated template not only semantically representative but also anatomically unbiased, which is crucial for following groupwise analysis.

\subsubsection{Exploration on loss design}
We propose $\mathcal{L}_{STAM}$ in InstantGroup to encourage the warped template to be close to the input. We explored adding the other similarity loss term that evaluates the pairwise alignment in the template space $\mathcal{L}_{pair} = \mathcal{L}_{sim}(I_1 \circ \phi_1^{-1}, I_2 \circ \phi_2^{-1})$. We add $\mathcal{L}_{pair}$ to the framework and present the comparison as (f), (g),  and (h). (h) achieves a high Dice Score but is less superior than (d) in Centrality and AvgDisp. 
As we compare (e) and (f), where the difference is to replace $\mathcal{L}_{STAM}$ by $\mathcal{L}_{pair}$, the proposed InstantGroup exhibits the overall better performance.

Secondly, we replaced $\mathcal{L}_{DIM}$ with its alternative (method e) $\mathcal{L}_{DIM-abl} = ||u_1||_2^2 + ||u_2||_2^2$ that minimizes the total magnitude of the displacement fields instead of encouraging them to be inverse. The results demonstrate the advantage of the proposed method over the variant.

\subsubsection{Hyperparameters on unbiasedness regularization}
Figure \ref{fig_opposite_param} presents the Dice Score, Centrality, and AvgDisp for the validation group set with varying unbiasedness regularization hyperparameters $\lambda_{dim}$. The proposed loss term $\mathcal{L}_{DIM}$ is designed to enforce the two displacement fields from the input images to the template to be inverse, thus encouraging the reconstructed middle template image to be at the center of the two inputs. 
When $\lambda=0$, it represents the Dual VAE without incorporating $\mathcal{L}_{DIM}$.
The Dice Score demonstrates an initial increase along the growth of $\lambda_3$, indicating that the regularization on unbiasedness can benefit the groupwise registration accuracy to a certain degree. When $\lambda$ surpasses a threshold, the Dice Score starts to decline. The Centrality and AvgDisp steadily decrease with a higher regularization hyperparameter on the unbiasedness term, which indicates better unbiasedness. Balancing the registration accuracy and unbiasedness leads to the selection of $\lambda_{dim}$ within [5, 7.5].
It could also be observed that the performance of DSC is quite tolerant with good results to the exploration of $\lambda_{DIM}$ adjusting the unbiasedness, which indicates few trade-offs and a harmonious balance between the metrics.

\begin{figure}[!t]
\centerline{\includegraphics[width=1\columnwidth]{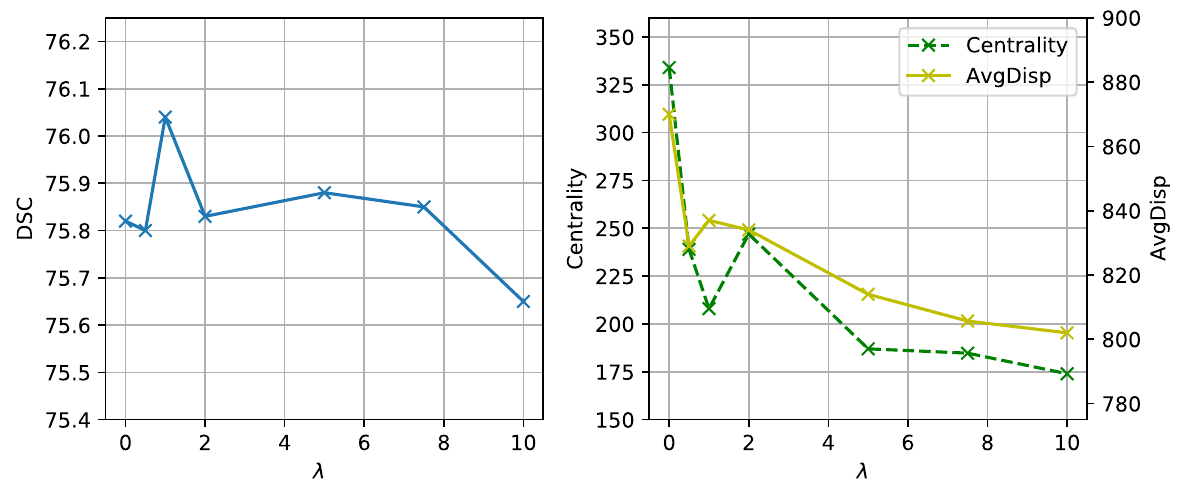}}
\vspace{-0.2cm}
\caption{Performance of InstantGroup on validation data with different hyperparameters $\lambda_{dim}$ for $\mathcal{L}_{DIM}$ for inverse displacement fields.}
\vspace{-0.2cm}
\label{fig_opposite_param}
\end{figure}

\subsubsection{Combined dataset training performance validation}
To validate the generalization of InstantGroup that whether a single model trained on the various datasets can generate custom templates successfully as the model trained specifically on each dataset. We trained a model on the combined dataset OASIS + ADNI and compared the performance of each dataset. The results are shown in Table \ref{combined}. For OASIS, the performance of the model trained on OASIS + ADNI drops a bit compared with that trained purely on OASIS. This could be due to the heterogeneity of the combined dataset, and the domain shift between OASIS and ADNI makes the learned latent space not as continuous and smooth as that trained on the specific dataset. Techniques like fine-tuning or domain adaptation \cite{guan2021domain} will further ensure robust performance across diverse cohorts.  
For ADNI, the performance of the model trained with the combined datasets is slightly improved. Therefore, for both datasets, the performance of the combined dataset training model maintains the state-of-the-art level. This illustrates the generalizability of the proposed InstantGroup and its potential to become a universal model enabling custom template generation for various datasets through multiple-dataset training.

\begin{table}[h]
\centering
\caption{Comparison of performance on OASIS and ADNI of models trained on combined datasets and separate datasets.}
\vspace{0.1cm}
\begin{tabular}{ c c c c c c}
\hline
\multicolumn{1}{c|}{\multirow{2}{*}{Method}} & \multicolumn{2}{c|}{Dataset}    & \multicolumn{1}{c|}{\multirow{2}{*}{Centrality$\downarrow$}} & \multicolumn{1}{c|}{\multirow{2}{*}{AvgDisp$\downarrow$}} & \multirow{2}{*}{DSC$\uparrow$} \\ \cline{2-3}
\multicolumn{1}{c|}{}                        & OASIS                             & \multicolumn{1}{c|}{ADNI}         & \multicolumn{1}{c|}{}                            & \multicolumn{1}{c|}{}                         &                      \\ \hline
\multicolumn{6}{c}{Performance evaluated on OASIS}                                                                                                                                                                                             \\ \hline
\multicolumn{1}{c|}{a}                       & \multicolumn{1}{c|}{$\checkmark$} & \multicolumn{1}{c|}{}             & \multicolumn{1}{c|}{\textbf{152.9}}              & \multicolumn{1}{c|}{\textbf{867.7}}           & \textbf{74.33}       \\ \hline
\multicolumn{1}{c|}{c}                       & \multicolumn{1}{c|}{$\checkmark$} & \multicolumn{1}{c|}{$\checkmark$} & \multicolumn{1}{c|}{161.2}                       & \multicolumn{1}{c|}{883.8}                    & 74.28                \\ \hline
\multicolumn{6}{c}{Performance evaluated on ADNI}                                                                                                                                                                                              \\ \hline
\multicolumn{1}{c|}{b}                       & \multicolumn{1}{c|}{}             & \multicolumn{1}{c|}{$\checkmark$} & \multicolumn{1}{c|}{163.7}                       & \multicolumn{1}{c|}{899.5}                    & \textbf{77.01}       \\ \hline
\multicolumn{1}{c|}{c}                       & \multicolumn{1}{c|}{$\checkmark$} & \multicolumn{1}{c|}{$\checkmark$} & \multicolumn{1}{c|}{\textbf{159.4}}              & \multicolumn{1}{c|}{\textbf{896.0}}           & 76.99                \\ \hline
\end{tabular}
\vspace{-0.2cm}

\label{combined}
\end{table}

\section{Discussion}
\subsection{Feature fusion strategy of InstantGroup}}
The efficiency and flexibility of InstantGroup are fundamentally rooted in its Dual VAE scheme and feature fusion strategy, which employs the averaging operation of latent representations. This enables the framework to effectively integrate features of arbitrary group sizes with low computational costs. InstantGroup ensures the reliable convex combination of the latent vectors by enforcing the continuity and smoothness of the latent space to generate a meaningful and comprehensive representation.
We explored alternative fusion strategies that are able to aggregate arbitrary lengths of information instantly and compared the average fusion strategy with a self-attention-based fusion strategy. The attention-like fusion computes the query $Q$ and key $K$ from the $N\times d$ variable ($N$ latent vectors of length $d$) by two MLPs. Then, the attention matrix is computed from $Q$ and $K$ as the weights of the vectors. This fusion turns to the weighted average of the latent representations. The comparison between the original average fusion and attention-based fusion is presented in Table \ref{fusion}. We can find that attention fusion is not obviously outperforming average fusion with limited decreases in Centrality while increasing the number of trainable parameters. The comparison indicates that average fusion is a feasible strategy for the framework because of its simplicity, reliability, and effectiveness.

\renewcommand{\arraystretch}{1.1}
\begin{table}[h]
\centering
\caption{Comparison on the feature fusion strategies. Average fusion refers to the fusion InstantGroup uses and Attention fusion is an alternative fusion strategy.}
\vspace{-0.2cm}
\begin{tabular}{c|c|c|c|c}
\toprule
Method & Centrality$\downarrow$ & AvgDisp$\downarrow$ & DSC$\uparrow$& $\#$Params (M) \\
\hline
Average fusion     & 152.9 & 867.7 & 74.33 &   1.3 \\
\hline
Attention fusion    & 147.2 & 870.8 & 74.23 & 2.1  \\
\bottomrule
\end{tabular}
\label{fusion}
\vspace{-0.2cm}
\end{table}

\subsection{On the selection of group-tailored templates over generic templates}
In this study, we advocate for the use of group-tailored templates in groupwise image registration, in contrast to relying on a single generic template constructed from a large population. From the theoretical perspective, the construction of a template that serves as the average of the group helps reduce inter-subject variability during the registration. In medical imaging, where population variations are affected by age, acquisition, disease, and other implicit factors, a generic template often fails to serve as a reliable reference for subtle structural analysis.  
We conducted experiments to quantitatively compare the performance of templates generated in both settings for small groups of four subjects in OASIS. In Table \ref{vs}, DeformT-1 refers to the original DeformT framework that generates a general template over the training set. DeformT-2 and InstantGroup are generating tailored templates for different groups. The results show that both tailor-based methods generate less biased templates with better deformation regularity and comparable registration accuracy compared with a generic template.  
\vspace{-0.2cm}

\renewcommand{\arraystretch}{1.1}
\begin{table}[h]
\centering
\caption{Performance comparison between generic templates and group-tailored templates. "Tailor" refers to the method that generates group-tailored templates ($\checkmark$) instead of general templates ($\times$).}
\vspace{-0.2cm}
\begin{tabular}{c|c|c|c|c|c|c}
\toprule
Method & Tailor & Central$\downarrow$ & AvgDisp$\downarrow$ & DSC$\uparrow$ & LogJ$_{95}$ & SDLogJ  \\
\hline
DeformT-1    & $\times$ & 631.5 & 996.3 & 75.31 & 0.0016 & 0.1575 \\
\hline
DeformT-2    & $\checkmark$ & 342.0 & 875.7 & 75.20 &  0.0015 & 0.1463 \\
\hline
InstantGrp  & $\checkmark$   & 284.0 & 812.3 & 75.66 & 0.0015 & 0.1354  \\
\bottomrule
\end{tabular}
\label{vs}
\vspace{-0.2cm}
\end{table}

\subsection{Extension and interaction with existing systems}
InstantGroup is proposed to efficiently generate group-tailored templates for various target groups. Therefore, the framework can be seamlessly integrated into existing groupwise image registration workflows, serving as an initial template generator. Based on the templates generated by InstantGroup, traditional iterative optimization methods can perform refined construction and registration. This kind of grafting can combine both speed and quality and, therefore, benefits the field of groupwise registration.

\section{Conclusion}
In this paper, we propose InstantGroup, a scalable and efficient groupwise template generation framework, the first learning-based method capable of handling groups of any size without requiring extra training. We introduce a Dual VAE backbone that processes pairs of inputs simultaneously using shared-weight networks with the Displacement Inversion Module (DIM) and Subject-Template Alignment Module (STAM).
DIM ensures anatomically centered and unbiased template generation by enforcing inverse displacement fields from each input to the template. 
STAM improves the image quality and reliability by measuring the similarity between the deformed template and each input scan.
Experimental results demonstrate that InstantGroup achieves superior unbiasedness and accuracy in template generation and drastically reduces computational time. 
Its flexibility and efficiency make it ideal for handling large-scale datasets. In conclusion, InstantGroup significantly advances GIR by providing a robust, efficient, and scalable solution for template generation.

% \appendices
\bibliographystyle{IEEEtran}
\bibliography{refs.bib}

\end{document}